\documentclass[useAMS,usenatbib]{mn2e}
\usepackage{url}
\font\japit = cmti10 at 10truept
\usepackage{epsfig}

\title
     [Angular Masks]
{\vglue-3.0truecm
\centerline{\japit Submitted to Monthly Notices}
\vglue 2.5truecm
\noindent
A scheme to deal accurately and efficiently with complex angular masks in galaxy surveys
\author[A. J. S. Hamilton and M. Tegmark]
     {A. J. S. Hamilton$^1$ and Max Tegmark$^2$ \\
	$^1$JILA and Dept.\ Astrophysical \& Planetary Sciences,
	Box 440, U. Colorado, Boulder CO 80309, USA; \\
	\ Andrew.Hamilton@colorado.edu; http:$/\!/$casa.colorado.edu/$\sim$ajsh/ \\
	$^2$Dept. of Physics, Univ. of Pennsylvania, Philadelphia, PA 19104, USA;
	max@physics.upenn.edu; http:$/\!/$www.hep.upenn.edu/$\sim$max/}
}

\newcommand{\bmi}{\bmath}

\newcommand{\dd}{{\rmn d}}	% MNRAS
\newcommand{\e}{{\rmn e}}	% MNRAS
\newcommand{\im}{{\rmn i}}	% MNRAS

\newcommand{\str}{{\rmn{str}}}

\newcommand{\el}{\ell}

\newcommand{\bL}{{\bmi L}}
\newcommand{\bn}{{\bmi n}}
\newcommand{\bx}{{\bmi x}}
\newcommand{\by}{{\bmi y}}
\newcommand{\bz}{{\bmi z}}

\newcommand{\aj}[2]{AJ, #1, #2}

\newcommand{\apj}[2]{ApJ, #1, #2}
\newcommand{\apjs}[2]{ApJS, #1, #2}

\newcommand{\mn}[2]{MNRAS, #1, #2}

\newcommand{\mangle}{{\sc mangle}}

\newcommand{\twoqzfig}{
    \begin{figure*}
    \begin{minipage}{175mm}
    \begin{center}
    \epsfxsize=6.5in
    \epsfbox{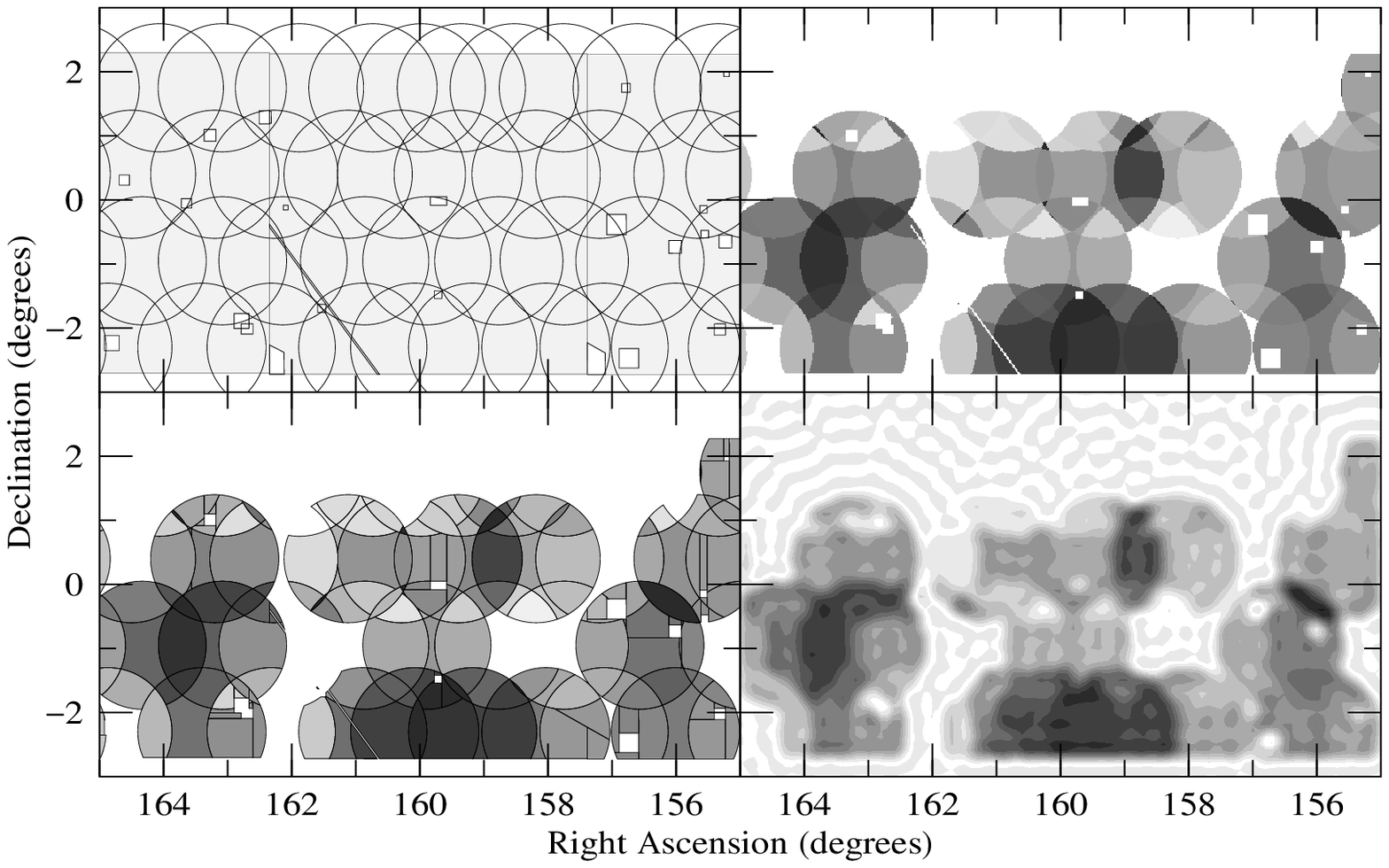}
    \end{center}
    \caption[1]{\small
    \label{twoqzfig}
A small piece of the northern angular mask of the 2QZ 10k release
(Croom et al.\ 2001).
(Top left) Delineation of the
$5^\circ \times 5^\circ$ UKST plates,
the holes in the UKST plates,
and the $2^\circ$ spectroscopic fields
of the 2QZ survey.
(Top right) The angular completeness of the survey,
in the $1^\prime \times 1^\prime$ pixelized form
provided by the 2QZ team.
(Bottom left) The 2QZ mask balkanized into polygons.
(Bottom right) The 2QZ mask reconstructed from harmonics
up to $\el = 1000$.
    }
    \end{minipage}
    \end{figure*}
}

\newcommand{\balkanizefig}{
    \begin{figure}
    \begin{center}
    \leavevmode
    \epsfxsize=1.8in
    \epsfbox{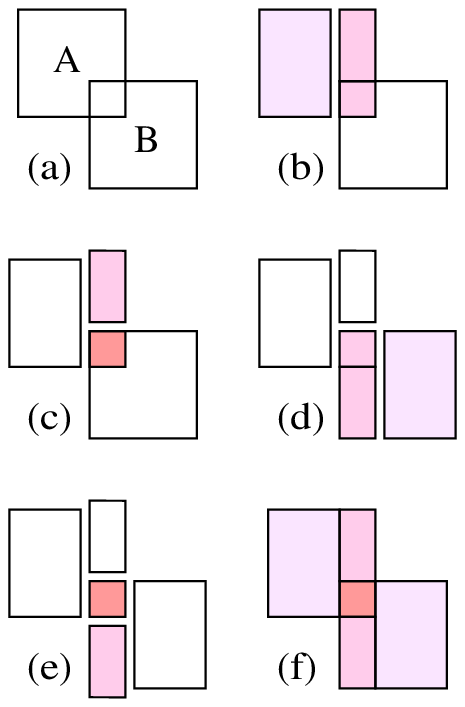}
    \end{center}
    \caption[1]{\small
    \label{balkanizefig}
An example of stage 1 of the balkanization procedure.
(a) The original system consists of two overlapping polygons,
A at upper left, B at lower right.
(b) First, polygon A is split against polygon B,
which takes two iterations.
In the first iteration,
A is split along an edge of B.
Only one of the two parts of the split A,
the right part,
overlaps B.
(c) In the second iteration,
the overlapping part of A, the right part,
is further split along another edge of B.
Again, only one of the two parts of the now doubly split A,
the lower right part,
overlaps B.
Since this lower right part is equal to the
intersection AB of A and B,
the process of splitting A against B terminates.
(d) Now, polygon B is split against polygon A.
Again, this takes 2 iterations.
In the first iteration,
B is split along an edge of A.
Only one of the two parts of the split B,
the left part,
overlaps A.
(e) In the second iteration,
the overlapping part of B, the left part,
is further split along another edge of A.
Again, only one of the two parts of the now doubly split B,
the upper left part,
overlaps A.
Since this upper left part is just equal to the intersection AB,
the process of splitting B against A terminates.
(f) The final system comprises 5 non-overlapping polygons,
consisting of 2 polygons inside A and outside B,
2 more polygons inside B and outside A,
and 1 polygon which is the intersection AB of A and B.
    }
    \end{figure}
}

\newcommand{\discpolfig}{
    \begin{figure}
    \begin{center}
    \leavevmode
    \epsfxsize=1in
    \epsfbox{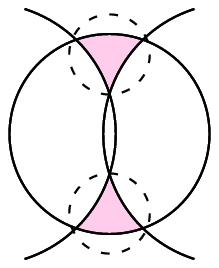}
    \end{center}
    \caption[1]{\small
    \label{discpolfig}
A polygon (shaded)
that contains two connected parts.
The polygon is the intersection of three caps,
the regions bounded by solid lines,
and therefore qualifies as a legitimate single polygon,
even though it contains two connected parts.
Stage 2 of the balkanization procedure
subdivides such disconnected polygons into connected parts
by computing the connected boundaries of the polygon,
and lassoing each connected boundary with an extra circle
(dashed lines).
    }
    \end{figure}
}

\newcommand{\eyefig}{
    \begin{figure}
    \begin{center}
    \leavevmode
    \epsfxsize=1in
    \epsfbox{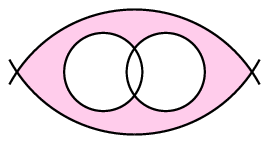}
    \end{center}
    \caption[1]{\small
    \label{eyefig}
A polygon (shaded)
that is connected but not simply-connected.
The polygon, which is the intersection of four caps,
has two distinct connected boundaries,
but the two boundaries belong to two separate groups,
and therefore do not need to be lassoed.
Stage 2 balkanization accepts the polygon as is.
    }
    \end{figure}
}

\newcommand{\discpolsfig}{
    \begin{figure}
    \begin{center}
    \leavevmode
    \epsfxsize=1in
    \epsfbox{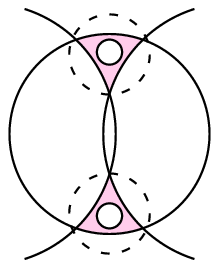}
    \end{center}
    \caption[1]{\small
    \label{discpolsfig}
A polygon (shaded)
similar to that of Figure~\protect\ref{discpolfig},
but with the addition of two circular holes.
The polygon has four distinct connected boundaries
belonging to three groups of circles.
Only the two boundaries in the same group need lassoing (dashed lines).
The two holes form separate groups, which do not need lassoing.
Stage 2 balkanization subdivides the polygon successfully into two parts.
    }
    \end{figure}
}

\newcommand{\fractalfig}{
    \begin{figure}
    \begin{center}
    \leavevmode
    \epsfxsize=2.5in
    \epsfbox{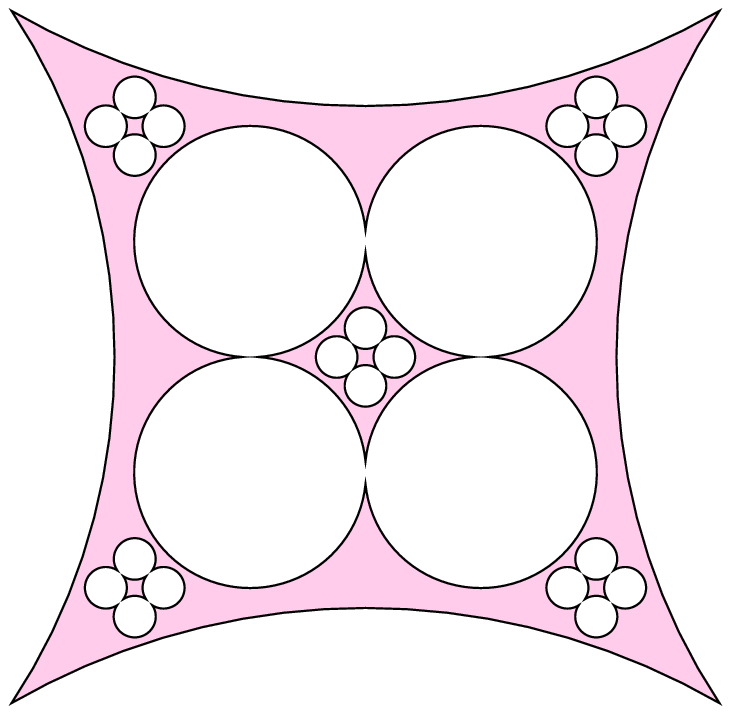}
    \end{center}
    \caption[1]{\small
    \label{fractalfig}
A multiply-connected polygon (shaded)
containing seven connected parts
bounded by thirteen distinct connected boundaries
belonging to seven groups of circles.
Stage 2 balkanization subdivides the polygon successfully into its seven parts.
    }
    \end{figure}
}

\newcommand{\eggboxfig}{
    \begin{figure}
    \begin{center}
    \leavevmode
    \epsfxsize=1in
    \epsfbox{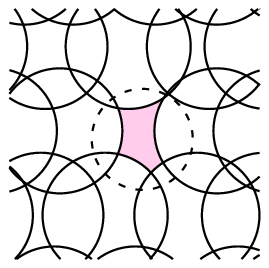}
    \end{center}
    \caption[1]{\small
    \label{eggboxfig}
A simply-connected polygon (shaded) with a large number of caps.
None of these caps can be discarded, since each excludes a small piece of sky.
Stage 2 balkanization lassos the polygon with an extra circle
(dashed line),
and, by incorporating the cap bounded by this circle,
is able to discard many of the original caps as superfluous.
    }
    \end{figure}
}

\newcommand{\chainfig}{
    \begin{figure}
    \begin{center}
    \leavevmode
    \epsfxsize=1in
    \epsfbox{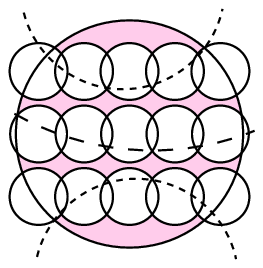}
    \end{center}
    \caption[1]{\small
    \label{chainfig}
A polygon (shaded) consisting of four connected parts
bounded by four boundaries that all belong to the same group.
Here stage 2 balkanization takes two passes to succeed.
In the first pass,
the top and bottom boundaries of the polygon are lassoed successfully
(dashed lines),
but the middle two boundaries cannot be lassoed with single circles
that fully enclose one boundary while fully excluding all other boundaries.
In the second pass,
the middle two boundaries are successfully partitioned
with a final lasso (short dashed line).
    }
    \end{figure}
}

\newcommand{\vfig}{
    \begin{figure}
    \begin{center}
    \leavevmode
    \epsfxsize=1in
    \epsfbox{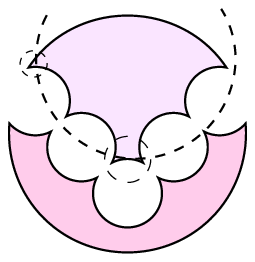}
    \end{center}
    \caption[1]{\small
    \label{vfig}
A polygon (shaded) consisting of two connected parts
bounded by two boundaries that belong to the same group.
Neither boundary can be lassoed with a circle
that fully encloses one boundary while fully excluding the other boundary.
Stage 2 balkanization splits the polygon into two
with a best-attempt lasso (dashed line)
that encloses as much as possible of one boundary (the upper boundary),
while fully excluding the other boundary (the lower boundary),
and then submits each of the two polygons to a second pass.
In the second pass,
the polygon enclosed by (above) the best-attempt lasso
is found to contain one boundary, and therefore needs no further splitting,
while the polygon outside (below) the lasso
is found to have three connected boundaries,
which are successfully partitioned with two more lassos (thin dashed lines).
Stage 2 balkanization thus successfully splits the original polygon
into a total of four non-overlapping parts.
    }
    \end{figure}
}

\newcommand{\wfig}{
    \begin{figure}
    \begin{center}
    \leavevmode
    \epsfxsize=2.5in
    \epsfbox{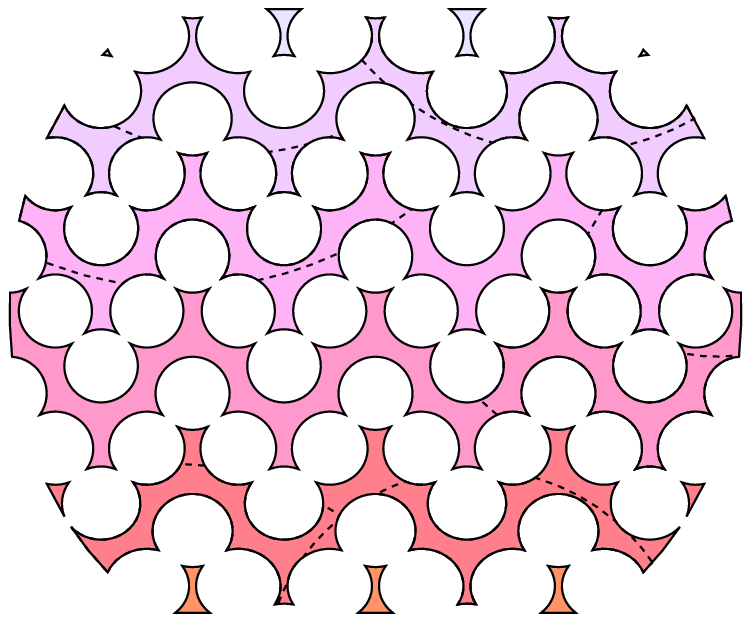}
    \end{center}
    \caption[1]{\small
    \label{wfig}
A tortuous polygon with parts designed to be difficult to lasso.
Stage 2 balkanization splits the polygon forcibly 8 times (dashed lines),
ultimately balkanizing the 13 connected parts of the polygon into 34 polygons.
    }
    \end{figure}
}

\newcommand{\multfig}{
    \begin{figure}
    \begin{center}
    \leavevmode
    \epsfxsize=3in
    \epsfbox{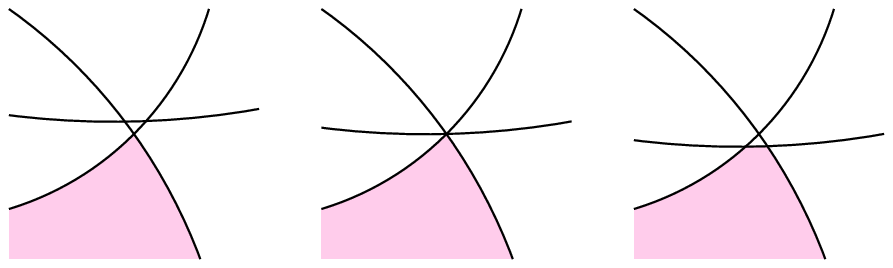}
    \end{center}
    \caption[1]{\small
    \label{multfig}
Example of how the topology near multiply-intersecting circles
can vary as a result of a tiny change in the location of the circles,
and can therefore be sensitive to numerical roundoff.
The horizontal edge crosses
in the left panel just above,
in the middle panel exactly at,
and in the right panel just below
the intersection of the diagonal edges.
The strategy to deal with the problem
is to test the topology of each vertex around a polygon for consistency,
as described in the text.
If an inconsistency is detected,
then the angular tolerance for considering nearly coincident intersections
to be exactly coincident is increased, until consistency is achieved.
    }
    \end{figure}
}

\newcommand{\kissfig}{
    \begin{figure}
    \begin{center}
    \leavevmode
    \epsfxsize=3in
    \epsfbox{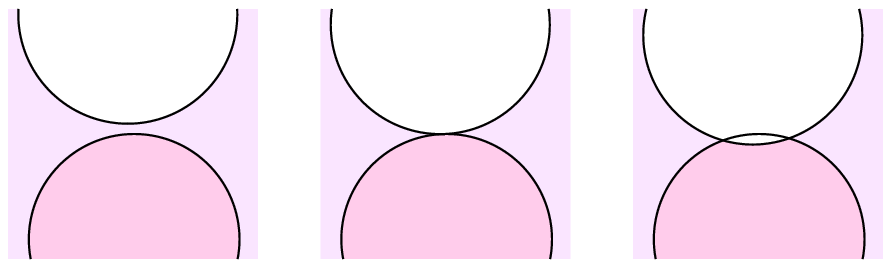}
    \end{center}
    \caption[1]{\small
    \label{kissfig}
Example of how the topology near kissing circles
can vary as a result of a tiny change in the location of the circles,
and can therefore be sensitive to numerical roundoff.
The strategy for dealing with near kissing circles
is similar to that for near multiply-intersecting circles.
    }
    \end{figure}
}

\newcommand{\flowerfig}{
    \begin{figure}
    \begin{center}
    \leavevmode
    \epsfxsize=2in
    \epsfbox{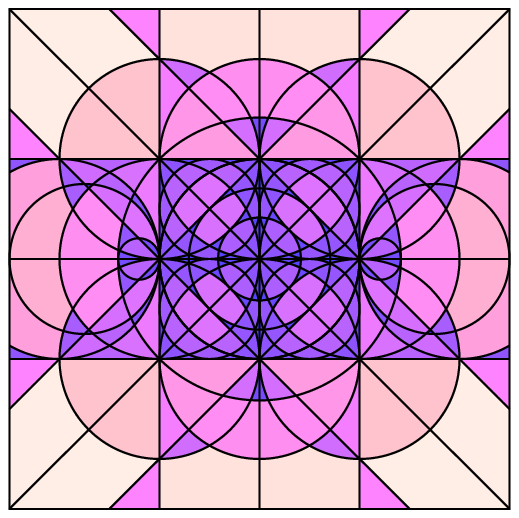}
    \end{center}
    \caption[1]{\small
    \label{flowerfig}
A `difficult' mask whose polygons
have many multiple and near multiple intersections,
and many kisses and near kisses.
Polygons are coloured according to their areas.
    }
    \end{figure}
}

\newcommand{\unifyfig}{
    \begin{figure}
    \begin{center}
    \leavevmode
    \epsfxsize=1.8in
    \epsfbox{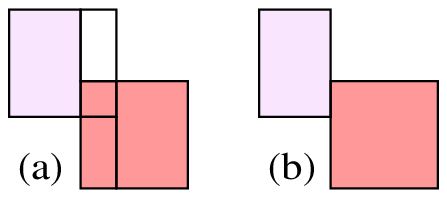}
    \end{center}
    \caption[1]{\small
    \label{unifyfig}
An example of the unification procedure.
(a) The initial balkanized, weighted mask
consists of 5 polygons as shown,
perhaps those from Figure~\protect\ref{balkanizefig}.
The upper middle polygon happens to have zero weight,
while the 3 lower polygons each happen to have the same non-zero weight.
(b)
Unification first removes the upper middle polygon,
which has zero weight.
Then, in a first pass,
unification merges the two lower of the middle polygons,
by removing the abutting edge between them.
In a second pass,
unification merges the lower right polygon
with the previously merged lower middle pair,
again by removing the abutting edge between them.
In a third pass,
unification finds no more adjacent pairs of polygons with the same weight
that can be merged by removing an abutting boundary,
and the procedure terminates,
leaving two polygons as shown.
    }
    \end{figure}
}

\newcommand{\nounifyfig}{
    \begin{figure}
    \begin{center}
    \leavevmode
    \epsfxsize=1in
    \epsfbox{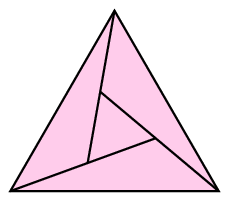}
    \end{center}
    \caption[1]{\small
    \label{nounifyfig}
A system of 4 polygons that
unification fails to merge into a single polygon,
because no pair of polygons can be merged. 
    }
    \end{figure}
}

\newcommand{\deftable}{
    \begin{table*}
    \begin{minipage}{175mm}
    \caption{Definitions of Terms, in Alphabetical Order}
    \label{deftable}
    \begin{tabular}{@{}lp{155mm}}
Term &
	Definition
\\
boundary &
	A set of edges bounding a polygon.
\\
cap &
	A spherical disk, a region above a circle on the unit sphere. 
\\
circle &
	A line of constant latitude with respect to some arbitrary polar axis on the unit sphere. 
\\
edge &
	An edge is part of a circle.  A polygon is enclosed by its edges. 
\\
great circle &
	A line of zero latitude with respect to some arbitrary polar axis
	on the unit sphere.
	A great circle is a circle,
	but a circle is not necessarily a great circle. 
\\
group &
	The circles of a polygon partition into groups:
	two circles are friends, belonging to the same group, if they intersect,
	(anywhere, not necessarily inside the polygon),
	and friends of friends are friends.
\\
mask &
	The union of an arbitrary number of weighted polygons. 
\\
polygon &
	The intersection of an arbitrary number of caps. 
\\
rectangle &
	A special kind of polygon, a rectangular polygon
	bounded by lines of constant longitude and latitude. 
\\
vertex &
	A point of intersection of two circles.
	A vertex of a polygon is a point where two of its edges meet. 
\\
weight &
	The weight assigned to a polygon.
	The spherical harmonics of a mask is the sum of the spherical harmonics
	of its polygons, each weighted according to its weight.
	A weight of 1 is the usual weight.
	A weight of 0 signifies an empty polygon, a hole.
	In general the weight may be some arbitrary positive or negative
	real number.
    \end{tabular}
    \end{minipage}
    \end{table*}
}

\begin{document}

\maketitle

\begin{abstract}
This paper presents
a scheme to deal accurately and efficiently with complex angular masks,
such as occur typically in galaxy surveys.
An angular mask is taken to be
an arbitrary union of arbitrarily weighted angular regions
bounded by arbitrary numbers of edges.
The restrictions on the mask are
(i) that each edge must be part of some circle on the sphere
(but not necessarily a great circle),
and
(ii) that the weight within each subregion of the mask must be constant. 
The scheme works by resolving a mask into disjoint polygons,
convex angular regions bounded by arbitrary numbers of edges.
The polygons may be regarded as the `pixels' of a mask,
with the feature that the pixels are allowed to take a rather general shape,
rather than following some predefined regular pattern. 
Among other things,
the scheme includes facilities to compute
the spherical harmonics of the angular mask,
and Data-Random $\langle DR \rangle$ and Random-Random $\langle RR \rangle$
angular integrals.
A software package \mangle\ which implements this scheme,
along with complete software documentation,
is available at
\url{http://casa.colorado.edu/~ajsh/mangle/}.
\end{abstract}


%\clearpage

\section{Introduction}
\label{intro}

With massive new data sets shrinking the statistical error bars on cosmological
quantities, it is becoming increasingly important to avoid inaccuracies in their modeling
and analysis. For example, 
one of the less interesting aspects of modern studies of large scale structure
is having to deal with complex angular masks.
This humble but essential task is rendered more time-consuming
by the fact that angular masks may require updating
%possibly frequently,
as a survey progresses.
The purpose of this paper is to describe a scheme
intended to remove much of the drudgery
and scope for inadvertent error or unnecessary approximation
involved in defining and using angular masks.
%so that one can get on to doing real science.
The scheme is implemented in a publically available software package
\mangle, which can be obtained, along with complete documentation, from
\url{http://casa.colorado.edu/~ajsh/mangle/}.
The present paper is not a software manual:
for that, visit the aforesaid website.
Rather, the purpose of this paper is
to set forward the philosophy and to detail the algorithms
upon which the software is built.

Angular masks of galaxy survey have grown progressively
more complicated through the years.
The first complete redshift surveys of galaxies,
the first Center for Astrophysics redshift survey
(CfA1; \citealt{Huchra}),
and the first Southern Sky Redshift Survey
(SSRS1; \citealt{daCosta}),
had rather simple angular boundaries,
defined by simple cuts in declination and in galactic latitude.

The level of complexity increased with the {\it IRAS\/} redshift surveys.
The first of these,
the {\it IRAS\/} 2~Jy redshift survey
\citep{Strauss},
in addition to a cut in galactic latitude,
excluded 1465 lunes of high cirrus or contamination by
Local Group galaxies,
each lune being an approximately $1^\circ \times 1^\circ$ square
with boundaries of constant ecliptic latitude and longitude.
The angular masks of subsequent {\it IRAS\/} surveys
followed a similar theme,
leading up to the PSCz survey
\citep{Saunders},
whose high-latitude mask
(the one most commonly used in large scale structure studies)
consisted of the whole sky less
11{,}477 $1^\circ \times 1^\circ$ ecliptic lunes.

The Automatic Plate Measuring (APM) survey
(\citealt{Maddox90a},b, 1996)
and associated surveys such as the APM-Stromlo survey
\citep{Loveday},
consisted of a union of a couple of hundred photographic plates,
cut to $5^\circ \times 5^\circ$,
each drilled with a sm\"org{\aa}sbord of holes to avoid bright stars,
satellite trails, plate defects, and the like.
The edges of the excluded holes were straight lines on photographic plates,
but unlike the {\it IRAS\/} surveys, the holes were not necessarily
rectangles, their boundaries were not necessarily lines of constant
latitude and longitude,
and different holes could overlap.

\twoqzfig

The Anglo-Australian Telescope 2 degree Field survey
(2dF; \citealt{Colless}; \citealt{Lewis})
is a redshift survey of galaxies from the APM survey,
and thus inherits the holey angular mask of the APM survey.
Superimposed on the APM backdrop,
the 2dF angular mask consists of several hundred overlapping $2^\circ$
diameter circular fields.
The various overlaps of the circular fields have,
at least in early releases of the data,
various degrees of completeness.

The Sloan Digital Sky Survey
(SDSS; \citealt{York})
has an angular mask comparable in complexity to that of the 2dF survey.
It consists of several stripes from the parent photometric survey,
peppered with holes masked out for a variety of reasons.
Superimposed on the stripes are circular fields from the redshift survey.
Recently,
the SDSS team used the \mangle\ scheme described in the present paper
as part of the business of computing the 3D galaxy power spectrum
\citep{Tegmark}.
With both the 2dF and SDSS data going public,
it has seemed sensible to publish the scheme
so that others can use it too.

The scheme described in the present paper began life
in the delightful atmosphere of an Aspen Center for Physics workshop in 1985.
The mathematics of harmonization (\S\ref{harmonize})
and other aspects of the computation of angular integrals
are written up in an Appendix to \citet{H93b}.
The methods described therein
were first applied by \citet{H93a},
and have been used regularly by him since that time.

The idea of adapting the methods to deal with angular masks in
a rather general way,
and in particular the concept of balkanization,
%(a word spontaneously invented by MT in the course of a telephone
%conversation with AJSH in late 2000),
is new to the present paper.
The \mangle\ software has been applied
to the 2dF 100k survey
by
\citet*{THX},
and to the SDSS
by
\citet{Tegmark}.

The figures in this paper
were prepared from files generated by the \mangle\ software.

\section{Mask definition}

Figure~\ref{twoqzfig}
shows a zoom of a small piece of the northern angular mask
of the 2dF QSO Redshift Survey (2QZ) 10k release
\citep{Croom}.
The angular mask of this survey
is defined by files
(downloadable from
\url{http://www.2dfquasar.org/Spec_Cat/masks.html})
giving the boundaries of:
(1) $5^\circ \times 5^\circ$ UKST plates,
(2) holes in UKST plates,
and (3) $2^\circ$ fields.
These boundaries are
illustrated in the top left panel of Figure~\ref{twoqzfig}.
The 2QZ team provide the completeness of the angular mask
in the $1^\prime \times 1^\prime$ pixelized form
illustrated in the top right panel of Figure~\ref{twoqzfig}.
The 2QZ mask is typical of the way that
angular masks are defined in modern galaxy surveys.

\deftable

Motivated by common practice,
an angular mask is defined in the present paper
to be an arbitrary union of arbitrarily weighted angular regions
bounded by arbitrary numbers of edges.
The restrictions on the mask are
\begin{enumerate}
\item
that each edge must be part of some circle on the sphere
(but not necessarily a great circle), and
\item
that the weight within each subregion of the mask must be constant. 
\end{enumerate}
This definition of an angular mask
by no means covers all theoretical possibilities,
but it does reflect the actual practices of makers of galaxy surveys.
%FIX
%and all current redshift surveys satisfy the above restrictions.
%FIX
The broad utility of spherical polygons to delineate angular regions
is widely appreciated; 
for instance, they play an integral part in the SDSS database.

The definition implies that an angular mask
is a union of arbitrarily weighted non-overlapping polygons.
A polygon is defined to be the intersection
of an arbitrary number of caps,
where a cap is defined to be a spherical disk,
a region on the unit sphere
above some line of constant latitude
with respect to some arbitrary polar axis.
For reference,
Table~\ref{deftable}
collects definitions of mask, polygon, cap,
and certain other terms used in this paper.

The bottom left panel of
Figure~\ref{twoqzfig}
shows the 2QZ mask `balkanized'
(see \S\ref{balkanize} below)
into non-overlapping polygons.
The bottom right panel of
Figure~\ref{twoqzfig}
shows the mask reconstructed from
spherical harmonics up to $\el = 1000$
(see \S\ref{harmonize} below).

\subsection{Polygon files}
\label{format}

The information specifying a mask
(its angular boundaries and completeness)
is collected in files which we refer to as `polygon files'.
Typically a command in the \mangle\ suite of software will:
\begin{enumerate}
\item
Read in one or more polygon files, possibly in different formats;
\item
Do something to or with the polygons;
\item
Write an output file, possibly a polygon file, or files.
\end{enumerate}

The strategy adopted in the \mangle\ software
is to permit the most flexible possible input format for polygon files,
the idea being to be able to read the files
provided by the makers of a galaxy survey
as far as possible in their original form, or perhaps mildly edited.
\mangle\ reads and writes several different formats of polygon files:
\begin{itemize}
\item
circle;
\item
vertices;
\item
edges;
\item
rectangle;
\item
polygon.
\end{itemize}

For convenience,
there are five additional formats
that provide useful information about polygons,
but that can only be written, not read,
because the information they provide is too limited,
or ambiguous, to specify polygons completely.
The five output only formats are:
area;
graphics;
id;
midpoint;
weight.

An abbreviated description of each format appears below;
see
\url{http://casa.colorado.edu/~ajsh/mangle/}
for full details.

%\subsubsection{Circle format}
%\label{circle}

The {\bf circle} format is able to describe polygons in all generality.
A circle is defined by the azimuth $\alpha$ and elevation $\beta$ of its north polar axis,
and by the angular radius, the polar angle $\theta$, of the circle.
Each circle defines a cap.
A polygon is an intersection of caps, and a line of the form 
\[
\alpha_1 \  \beta_1 \  \theta_1 \  \ldots \  \alpha_n \  \beta_n \  \theta_n
\]
containing $3n$ angles
defines a polygon with $n$ caps.

%\subsubsection{Vertices format}
%\label{vertices}

The {\bf vertices} format specifies polygons by a sequence of vertices,
assumed to be joined by great circles.
The general form of a line specifying a polygon in vertices format is 
\[
\alpha_1 \  \beta_1 \  \ldots \  \alpha_n \  \beta_n
\]
which defines a polygon with $n$ caps.
In vertices format, a line with $2n$ angles defines a polygon with n caps. 

%\subsubsection{Edges format}
%\label{edges}

The {\bf edges} format is a souped-up version of the vertices format.
Whereas the vertices format joins each pair of vertices
with a great circle, the edges format uses an additional point
(or additional points) between each pair of vertices
to define the shape of the circle joining the vertices.
Although the edges format retains more information about a polygon
than the vertices format,
in general it does not retain all information about a polygon.

%\subsubsection{Rectangle format}
%\label{rectangle}

A rectangle is a special kind of 4-cap polygon
bounded by lines of constant azimuth and elevation.
The {\bf rectangle} format is offered not only because some masks are defined this way
(for example, the {\it IRAS\/} masks),
but also because the symmetry of rectangles permits accelerated
computation of their spherical harmonics
(\S\ref{harmonize}).
A line in rectangle format looks like 
\[
\alpha_{\min} \  \alpha_{\max} \  \beta_{\min} \  \beta_{\max}
\]
with precisely 4 angles.

%\subsubsection{Polygon format}
%\label{polygon}

The {\bf polygon} format is the default output format for polygon files.
Besides the circle format,
it is the only other format that is able to describe
polygons in all generality without loss of information.
It stores each cap,
not as three angles as in the circle format,
but rather as a unit vector along the north pole of the cap,
together with a quantity $1 - \cos\theta$,
which is equal both to the area of the cap divided by $2\upi$,
and to half the square of the 3-dimensional distance
between the north pole and the cap boundary.
It seems doubtful that one would want to create an original mask file
in polygon format, since it is a bit peculiar,
%(and we do not pause to describe it here),
but it is the format used internally by the \mangle\ software,
and it specifies polygons in the manner expected for many years past
by the fortran backend.
%If you don't like that, blame history, which can be blamed for a lot of things. 
The advantage of the format is that
some computational operations are simpler and faster
if the cap axis is stored as a unit vector
rather than as an azimuth and elevation.

%\subsubsection{Graphics format}
%\label{graphics}

Of the purely output formats,
one of the most useful is the {\bf graphics} format,
which is useful for making plots of polygons.
The \mangle\ software does not incorporate any plotting software:
it is assumed that you have your own favourite plotting package. 
The graphics format is similar to edges format,
but is generally more economical.
Whereas in edges format there is a specified number of points per edge,
in graphics format there is a specified number of points per $2\upi$
of azimuthal angle along each edge.
Thus in graphics format curvier edges get more points than straighter edges. 
The graphics format is implemented only as output, not as input,
because of ambiguity in the interpretation of the format.

Another useful output format is the {\bf midpoint} format,
which returns a list containing the angular position of a point
inside each polygon of a mask.
This can be helpful in assigning weights to the polygons of a mask,
if you have your own software that returns a weight
given an angular position.
See \S\ref{midpoint} for more about how midpoints of polygons are computed.

The {\bf area}, {\bf id}, and {\bf weight} output formats
give lists of, respectively,
the areas, identity numbers, and weights (completenesses)
of the polygons of a mask.

\section{Resolving a mask into non-overlapping polygons}

One of the basic tasks that the \mangle\ software does
is resolve a mask into a set of non-overlapping polygons.
This takes place in a sequence of four steps,
elaborated in the subsections following.
\begin{enumerate}
\item
Snap;
\item
Balkanize;
\item
Weight;
\item
Unify.
\end{enumerate}

Resolving a mask into non-overlapping polygons
greatly simplifies the logic of dealing with a mask,
since it allows subsequent processing
(generation of random catalogues,
computation of spherical harmonics,
plotting, etc.)
to proceed without recourse to
the intricate hierarchy of overlapping geometric entities
and the associated complicated series of inclusion and exclusion rules 
that tend to characterize a survey mask.

Each individual polygon is by definition an intersection of caps.
Geometrically, this implies that a polygon is convex:
%(although not necessarily connected or simply-connected):
the interior angles at the vertices of a polygon are all less than $\upi$.
The requirement that a polygon be an intersection of caps
greatly simplifies the logic,
since it means that a point lies inside a polygon if and only if
it lies inside each of the caps of the polygon.

\subsection{Snap}
\label{snap}

The first thing that must be done on all the original polygon files
of a mask is to `snap' them.
This process identifies almost coincident cap boundaries
and snaps them together. 

The problem is that the positions of the intersections of two almost but
not exactly coincident circles (cap boundaries) on the unit sphere may be
subject to significant numerical uncertainty.
To avoid numerical problems,
such circles must be made exactly coincident.
You might think that that near-but-not-exactly-coincident circles would
hardly ever happen, but in practice they occur often, because a mask designer
tries to make two polygons abut,
but imprecision or numerical roundoff defeats an exact abutment. 

The snap process adjusts the edges of each polygon,
but it leaves the number and order of polygons the same as in the input file(s).
Edges that appear later in the input file(s) are snapped to earlier edges. 

The snap process offers four tunable tolerances:
\begin{itemize}
\item
Axis tolerance.
Are the axes of two caps within this angular tolerance of each other,
either parallel or anti-parallel?
If so, change the axis of the second cap to equal that of the first cap. 
\item
Latitude tolerance.
If the axes of two caps coincide, are their latitude boundaries within this tolerance of each other?
If so, change the latitude of the second cap to equal that of the first cap.
The two caps may lie either on the same or on opposite sides of the latitude boundary. 
\item
Edge tolerance, and edge to length tolerance.
Are the two endpoints and midpoint of an edge closer to a cap boundary than the lesser of
(a) the edge tolerance,
and (b) the edge to length tolerance
times the length of the edge?
In addition, does at least one of the two endpoints or midpoint of the edge
lie inside all other caps of the polygon that owns the cap boundary?
If so, change the edge to align with the cap boundary. 
\end{itemize}

The purpose of the first two of these tolerances,
the axis tolerance and the latitude tolerance, is obvious.
The remaining two tolerances,
the edge tolerance and the edge to length tolerance,
are necessary because it is possible for two edges,
if they are short enough, to almost coincide
even though the axes and latitudes of their corresponding caps
differ significantly.

By default, the three angular tolerances
(axis, latitude, and edge)
are all two arcseconds,
which is probably sufficient for typical large scale structure masks.
The tolerances can be tightened considerably before numerical problems
begin to occur, so it is fine to tighten the tolerance for a mask
whose edges are more precisely defined.
The default edge to length tolerance $0.01$,
should be fine in virtually all cases.

The snap process accomplishes its work in two stages: 
\begin{itemize}
\item
Snap axes and latitudes of pairs of caps together,
passing repeatedly through all pairs of caps until no more caps are snapped. 
\item
Snap edges of polygons to other edges,
again passing repeatedly through all pairs of caps until no more caps are snapped. 
\end{itemize}

As a finishing touch,
snap prunes each of the snapped polygons in order to eliminate superfluous caps,
those whose removal leaves the area of the polygon unchanged.

\subsection{Balkanize}
\label{balkanize}

The process of resolving a mask into disjoint polygons we dub `balkanization',
since it fragments an input set of possibly overlapping polygons
into many non-overlapping connected polygons.
The process involves two successive stages:
\begin{enumerate}
\item
Fragment the polygons into non-overlapping polygons,
some of which may be disconnected.
\item
Identify disconnected polygons and subdivide them into connected parts.
\end{enumerate}

\subsubsection{Balkanization Stage 1}
\label{balkanize1}

\balkanizefig

The algorithm for the first stage of balkanization is simple and pretty:
\begin{itemize}
\item[(a)]
Is the intersection of two polygons
neither empty nor equal to the first polygon?
If so, find a circle, a cap boundary, of the second polygon
that divides the first polygon,
and split the first polygon into two along that circle.
\item[(b)]
Iterate.
\end{itemize}
Notice that only one of the two parts of the split polygon
overlaps the second polygon,
and that only the overlapping part needs iterating.
For any pair of polygons,
iteration ceases when the overlapping part lies entirely inside the second
polygon.
The final overlapping part is equal to the intersection of the original
first polygon with the second polygon.
All other fragments of the first polygon lie outside the second polygon.

Figure~\ref{balkanizefig}
illustrates an example of the first stage of balkanization
for two overlapping polygons A and B.
First, A is split against B, which takes two iterations of the above cycle.
Then, B is split against A.
Again, this takes two iterations of the above cycle.
The final system consists of 5 non-overlapping polygons.

Note that splitting the system
shown in panel (a) of Figure~\ref{balkanizefig}
into its three connected parts
(the part of A that does not intersect B,
the part of B that does not intersect A,
and the intersection AB of A and B)
would not constitute a successful balkanization,
since two of these regions are not convex and hence not polygons.

One might ask,
why not stop at panel (c) in Figure~\ref{balkanizefig}?
Do not the three polygons there already form a satisfactory set of
non-overlapping polygons?
The answer is that
the intersection polygon AB may well have a weight different
from those of the non-overlapping parts of the parent A and B polygons
(this is typically true for example in the 2dF and SDSS surveys).
To deal with this eventuality,
balkanization must continue to completion, as illustrated in panel (e).

The question of whether two polygons overlap is
determined by computing the area of the intersection of the polygons.
The area is proportional to the monopole harmonic,
computed as described in \S\ref{harmonize}.
The intersection of two polygons is itself a polygon,
consisting of the intersection of the two sets of caps
defining the polygons.

\subsubsection{Balkanization Stage 2}
\label{balkanize2}

\discpolfig

Stage 1 of the balkanization procedure yields polygons
that can contain two or more connected parts, as illustrated
in Figure~\ref{discpolfig}.
Stage 2 attempts to subdivide such disconnected polygons into connected parts
by computing the connected boundaries of the polygon,
and lassoing (see \S\ref{lasso})
each connected boundary with an extra circle.

\eyefig

Figure~\ref{eyefig}
illustrates a polygon that has two distinct connected boundaries
by virtue of being not simply-connected rather than not connected.
A region is said to be {\bf simply-connected} if,
according to the usual mathematical definition,
it is connected and
any closed curve within it can be continuously shrunk to a single point.
Loosely speaking, this means that a simply-connected polygon has no holes.
%This property is independent of whether the region is {\bf connected},
%i.e., allows any two points within it to be connected by a curve within it.
%A polygon can be simply-connected but not connected,
%as in Figure~\ref{discpolfig},
%and connected but not simply-connnected,
%as in Figure~\ref{eyefig}.
Because it is connected,
the polygon of Figure~\ref{eyefig} need not be split.

The strategy to deal with non-simply-connected polygons
is based on the following theorem, proven in the Appendix:
{\bf A connected part of a polygon is simply-connected
if and only if all the boundaries of the connected part belong to a single group}.
A group\footnote{
Mathematically, the things here called groups here are equivalence classes,
not mathematical groups.
}
is defined here as follows:
two circles are friends, belonging to the same group, if they intersect
(anywhere, not necessarily inside the polygon),
and friends of friends are friends.
According to this definition of group, the circles
on a single connected boundary necessarily all belong to the same group.
However, the circles on two distinct connected boundaries
may or may not belong to the same group.

This theorem implies that
it is necessary to lasso only those boundaries of a polygon
that belong to the same group.
In Figure~\ref{discpolfig},
for example,
the two boundaries of the polygon
belong to the same group of three intersecting circles,
so these two boundaries must be lassoed,
partitioning the polygon into two parts.
In Figure~\ref{eyefig},
on the other hand,
the two boundaries
belong to two separate groups,
and need not be lassoed.

\discpolsfig

Figure~\ref{discpolsfig}
illustrates a more complicated polygon,
similar to the polygon of Figure~\ref{discpolfig}
but pierced with two circular holes.
The polygon contains four boundaries
belonging to three groups.
The two original boundaries inherited from Figure~\ref{discpolfig}
belong to the same intersecting group of circles,
but the additional two holes form two separate groups.
Here only the two boundaries belonging to the same group need lassoing.

\fractalfig

Figure~\ref{fractalfig}
illustrates a yet more complicated multiply-connected polygon.
The \mangle\ software balkanizes this polygon correctly into seven polygons,
a stringent test of the algorithms.

A corollary of the theorem proven in the Appendix
is that the polygon formed by the intersection of the caps bounded by
the circles of a single group must be a union of simply-connected parts.
For example,
the two parts of the polygon in Figure~\ref{discpolfig}
must be simply-connected -- which evidently they are --
because the circles of the polygon all belong to the same group.

In the course of the proof in the Appendix, it is shown that
if the boundary of a polygon falls into two (or more) groups,
then the circles of a second group must lie entirely inside
exactly one of the simply-connected parts of the polygon
bounded by the first group.
For example,
each of the two circles bounding the two holes in the polygon of
Figure~\ref{discpolsfig}
must lie entirely inside exactly one of the two simply-connected parts
of the original polygon from Figure~\ref{discpolfig},
which again is evidently true.

It follows from the statement of the previous paragraph
that in lassoing the connected boundaries of a group,
it is necessary to consider only the boundaries belong to the same group:
any boundary belonging to another group can be ignored,
because it must lie entirely inside one of the simply-connected parts
bounded by the first group.
Thus each lassoing circle is required to enclose fully its connected boundary,
while excluding fully all other connected boundaries
belonging to the same group;
there is no constraint on the lasso from boundaries belonging to other groups.

\eggboxfig

If a group of circles of a polygon defines a single boundary,
then that boundary needs no lassoing,
but stage 2 balkanization nevertheless attempts to lasso the boundary
if the number of caps of the group exceeds the number of vertices.
For example, in its original configuration the polygon
shown in Figure~\ref{eggboxfig}
has a large number of caps.
None of the caps can be discarded,
since each excludes a small piece of the sky.
Here it is advantageous to lasso the polygon with an extra circle,
allowing most of the original caps to be discarded as superfluous.

A lasso that lassos the lone boundary of a group
is discarded if the lasso completely encloses all the circles of the group
to which the boundary belongs.
For then either the lasso completely encloses the polygon,
in which case it is superfluous,
or else the lasso lies completely inside the simply-connected region
bounded by the lone boundary,
in which case the lasso, if kept,
would divide the simply-connected region in two, which would be incorrect.
If on the other hand the lasso of a lone boundary of a group
intersects at least one of the circles of the group,
then the lasso must completely enclose the simply-connected region
bounded by the lone boundary,
as in Figure~\ref{eggboxfig};
the lasso cannot lie inside the simply-connected region
because it is being assumed that the lasso intersects a circle of the group,
whereas no such circle can exist within the simply-connected region.

\chainfig

Stage 2 balkanization may need more than one pass to succeed.
Figure~\ref{chainfig}
shows an example of a polygon,
bounded by one large cap punctuated by fifteen small caps,
that contains four parts bounded by four boundaries
all belonging to the same single group.
The top and bottom boundaries can be lassoed successfully with single circles,
but the middle two boundaries cannot:
any circle that encloses either of the middle boundaries
necessarily intersects another boundary somewhere.
Here stage 2 balkanization succeeds by submitting the polygon to two passes.
In the first pass, the polygon is split into three polygons,
consisting of the top and bottom connected parts,
plus a third polygon containing the two middle parts.
In the second pass, the third polygon is split into two,
completing the partitioning of the original polygon into its four parts.

\vfig

In certain convoluted cases,
such as the polygon shown in Figure~\ref{vfig},
it can be impossible to lasso any of the connected boundaries
of the polygon with a circle that wholly encloses a connected boundary
while wholly excluding all other connected boundaries in the same group.
Stage 2 balkanization
gives up attempting to lasso a boundary after a certain maximum
number of attempts,
but it keeps a record of the best-attempt lasso,
the one that encloses as much as possible
of a boundary while wholly excluding all other boundaries in the same group.
Stage 2 balkanization proceeds to
split the polygon into two parts with the best-attempt lasso,
and then submits the two parts to a further pass.
The polygon of
Figure~\ref{vfig},
for example,
contains two parts bounded by two connected boundaries
neither of which can be lassoed with a circle that completely
encloses the boundary while completely excluding the other boundary.
Finding no satisfactory lasso,
stage 2 balkanization
splits the polygon into two polygons with a best-attempt lasso,
shown as a dashed line in Figure~\ref{vfig}.
The two polygons are then submitted to further passes of
stage 2 balkanization,
which in this case succeeds with one pass.
The upshot is that the original polygon is balkanized into four polygons.

\wfig

It is conceivable that the algorithm of the above paragraph could
continue for ever,
continually splitting a polygon into two and continually failing
to lasso successfully all the boundaries of the split polygons.
However, polygons that defy lassoing have to be filamentary in character
(long, thin, and windy),
such as that shown in Figure~\ref{wfig},
and splitting such a polygon in two
generally makes it less filamentary, like putting spaghetti in a blender.
In the case of the polygon of Figure~\ref{wfig},
stage 2 balkanization forcibly splits the polygon 8 times,
eventually balkanizing the 13-part polygon into 34 disjoint parts.
Suffice to say that we know of no polygon that fails the algorithm,
and it is possible that no such polygon exists.
If the reader finds one, please tell us about it.

In practice,
the \mangle\ software
bails out if a polygon has to be split forcibly in two
more than a certain maximum number (100) of times.
Even in this last gasp case,
the set of polygons output by balkanization
still constitutes a valid set of non-overlapping polygons
that completely tile the mask.
The only problem is that the `failed' polygons,
those which could not be partitioned completely,
may contain two or more disjoint parts with different weights.

%If stage 2 balkanization fails to partition a polygon successfully,
%it is not a disaster.
%The set of polygons output by balkanization
%still constitutes a valid set of non-overlapping polygons
%that completely tile the mask.
%The only problem is that the `failed' polygons
%may contain two or more connected parts
%with different weights.
%This problem should occur rarely
%(it has never occurred in the 2dF or SDSS masks),
%and one has to work hard to devise a polygon that
%is sufficiently convoluted to fail.
%So far we have not been able to invent a computer algorithm
%that would work in all possible cases,
%but a human should in most cases be able to find simple workaround.
%For example,
%a human might consider modifying the input polygons
%so as to avoid abstrusely complicated output polygons.

To finish,
balkanization prunes each of the balkanized polygons
in order to eliminate superfluous caps,
those whose removal leaves the area of the polygon unchanged.
For example,
pruning discards the many superfluous caps
of the polygon of Figure~\ref{eggboxfig}.
The caps are tested in order,
with any new lassoing cap being tested last,
so that the many superfluous caps are discarded,
and the lassoing cap is kept.

\subsubsection{Lasso}
\label{lasso}

The algorithm to lasso a connected boundary of a polygon
is to pick a point,
initially taken to be the barycentre of the
centres of the edges of the connected boundary
(or, if the connected boundary consists of a single circle,
the centre of that circle),
and find the circle centred on that point
which most tightly encloses the boundary.
The lassoing circle is enlarged slightly if possible,
as a precaution against numerical problems that might potentially occur
if the lasso just touched an edge or vertex of the boundary.

A lasso that lassos, i.e. that encloses completely,
one connected boundary of a polygon,
is required to exclude completely all other connected boundaries
belonging to the same group.
A lasso attempt can sometimes fail if a polygon has two or more
connected boundaries\footnote{
Actually a lasso attempt can in some circumstances fail
if a polygon has just one connected boundary,
if the centre point of the lasso is chosen in a maximally stupid fashion.
For example, if the connected boundary is a single great circle,
and if the centre of the lasso is stupidly chosen to lie on that great circle,
then the lasso will fail.
The problem is trivial to guard against.
}.
A lasso attempt fails if
the angular distance from the centre of the lasso
to the farthest point $\bx$
on the to-be-lassoed connected boundary
is greater than
the angular distance from the centre of the lasso
to the nearest point $\by$
on all other connected boundaries belonging to the same group.
If a lasso attempt fails,
then the centre point of the lasso is shifted over the unit sphere
along the vector direction from $\by$ to $\bx$,
by an amount that puts the centre point just slightly closer to
$\bx$ than $\by$.
The lasso is then reattempted.
The process of shifting the centre point and retrying a lasso
is repeated until either the lasso succeeds,
or until a certain maximum number of attempts has been made.

%If a pass of stage 2 balkanization of a polygon is partially successful,
%that is,
%if it succeeds in partitioning some but not all of the
%parts of a polygon,
%then the partitioned parts are added to the store
%of successfully partitioned polygons,
%and a new polygon is constructed from the remaining unpartitioned parts
%of the polygon,
%that is, from the intersection of the original polygon
%with the complement of the successful lassos.
%The new polygon is then subjected to a further
%pass of stage 2 balkanization.
%The polygon in Figure~\ref{chainfig} is an example of this process.
%Passes are repeated until
%either the polygon is fully partioned,
%or none of the connected boundaries of the polygon can be lassoed.
%Thus stage 2 balkanization fails on a polygon
%only if the polygon has two or more connected boundaries belonging to
%the same group,
%and none of the connected boundaries in the group can be lassoed.

\subsubsection{Multiply-intersecting and kissing circles}

\multfig

Multiple intersections occur where 3 or more circles (cap boundaries)
intersect at a single point.
Multiple intersections pose a potential source of numerical problems,
%partly because numerical roundoff can cause multiply-intersecting
%circles to fail to intersect at exactly the same point,
%and partly
because the topology around multiple intersections
may vary depending on numerics,
as illustrated in Figure~\ref{multfig}.

\kissfig

Circles kiss if they just touch.
Again, kissing circles pose a potential source of numerical problems,
because whether two circles kiss
may vary depending on numerics,
as illustrated in Figure~\ref{kissfig}.

\mangle\ is equipped to deal with both multiply-intersecting and kissing
circles, and should cope in almost all cases,
although it is possible to fool \mangle\
with a sufficiently complicated polygon,
for example a polygon whose vertices have a fractal distribution of separations.

The strategy is as follows.
Circles are considered to be multiply intersecting,
crossing at a single vertex,
if the intersections are closer than a certain tolerance angle.
Similarly, circles are considered to kiss,
touching at a single vertex,
if their kissing distance is closer than the tolerance angle.
The algorithm is friends-of-friends:
two vertices closer than the tolerance are friends,
and a friend of a friend is a friend.
The position of each vertex $ij$ of a polygon,
where edge $i$ intersects (or kisses) edge $j$,
is computed two different ways,
first as the intersection of edge $i$ with edge $j$,
then as the intersection of edge $j$ with edge $i$.
For each of the two ways of computing it,
the intersection is tested against all other circles,
to determine whether the intersection is or is not a vertex of the polygon,
that is, whether the intersection lies on the edge of the polygon,
or outside the polygon.
For consistency, the test should give the same result in both computations:
the intersection should be a vertex in both cases,
or it should not be a vertex in both cases.
If an inconsistency is detected, then the tolerance angle is doubled
(or set to a tiny number, if the tolerance is zero),
and the computation is repeated for the inconsistent polygon,
until consistency is achieved.
By default, the initial tolerance angle for multiple intersections
and kissings is $10^{-5}$ arcseconds.

In Figure~\ref{multfig},
the intersection of the two diagonals
is a vertex of the polygon in the left panel,
but is not a vertex in the right panel,
because it lies outside the polygon.
In the case of an exact multiple intersection,
as in the middle panel of Figure~\ref{multfig},
the intersection $ij$ is considered to be a vertex of the polygon
only if $i$ and $j$ are both edges of the polygon.
Thus the intersection of the two diagonals in the middle panel
of Figure~\ref{multfig} is a vertex, because both diagonals are edges,
but the intersection of the horizontal with either diagonal
is not a vertex, because the horizontal is not an edge of the polygon.
If $i$ is an edge,
and it intersects multiply with a bunch of other circles,
then the adjacent edge $j$ is formed by the circle
which `bends most tightly' around the polygon,
that is,
the circle whose interior angle at the vertex is the smallest,
or, if two circles subtend the same interior angle (within the tolerance angle),
then the circle whose polar angle $\theta$ is the smallest.

In Figure~\ref{kissfig},
the two circles $i$ and $j$ intersect
at no vertices in the left panel,
and at two vertices in the right panel.
In the case of an exact kiss,
as in the middle panel of Figure~\ref{kissfig},
the kissing point $ij$ is considered to be a vertex
only if $i$ and $j$ are both edges of the polygon.
Thus the kissing point $ij$ is {\em not\/} a vertex of the
the upper and lower polygons, the two disks,
but it is a vertex of the middle polygon, the pointy one.

Two polygons that just touch or kiss at a single isolated point
(or at a set of isolated points)
are considered to be disconnected from each other.
Thus for example the top and bottom polygons
in the middle panel of Figure~\ref{kissfig}
are considered to be disconnected from each other;
and similarly the left and right polygons
in the middle panel of Figure~\ref{kissfig}
are considered to be disconnected from each other.
%In this respect the definition of connected in this paper
%deviates from the usual mathematical definition.

In practice, consistency of the topology of the distribution of vertices around
a polygon is checked by means of a 64-bit check number.
If the intersection $ij$ of edge $i$ with edge $j$
(two edges can intersect at two separate points,
so $ij$ is an ordered pair,
going from edge $i$ to edge $j$
right-handedly around the boundary of the polygon)
is determined to be a vertex of the polygon,
then a 64-bit pseudo-random integer is added to the check number,
and if the same intersection $ij$ of edge $j$ with edge $i$
is determined to be a vertex of the polygon,
then the same 64-bit pseudo-random integer is subtracted from the check number.
For consistency, the check number should be zero for the entire polygon. 
It is conceivable, with probability 1 in $2^{64}$,
or less than 1 in 10 billion billion,
that the check number could evaluate to zero accidentally,
but this seems small enough not to worry about,
especially since inconsistency should be a rare occurrence in the first place.

\flowerfig

Figure~\ref{flowerfig}
illustrates a mask designed to be as `difficult' as possible:
it contains many multiply intersecting and nearly multiply-intersecting circles,
and many kissing and nearly kissing circles,
including several simultaneously
multiply-intersecting and multiply-kissing circles.
The \mangle\ software copes with this,
a non-trivial accomplishment.

The sum of the areas of the 332 polygons of the balkanized mask
of Figure~\ref{flowerfig}
differ from the area, $0.00761302 \ \str$,
of the overall bounding rectangle by $7 \times 10^{-14} \,\str$,
which is definitely satisfactory.
Given the algorithms,
one could expect the numerical uncertainty in the area
of a single polygon to be no better than machine precision times
$2\upi \ \str$, which on the machine used
for this computation was about $10^{-15} \, \str$.

\subsubsection{Order of polygons to be balkanized}

When balkanizing,
does the order of the polygons in the input polygon files matter?
The answer is yes,
if the input polygons overlap,
and if the overlapping polygons carry different weights.
As described in the following subsection~\ref{weight},
if two polygons overlap,
then the weight of the polygon that appears later in the input file(s)
overrides the weight of the earlier polygon.

If all polygons have the same weight (say 1),
then the order of the input polygon files does not really matter.
However, it may lead to a slightly smaller eventual polygon file
(after unifying, see \S\ref{unify})
if large, coarse polygons are put first,
and small, finely detailed polygons are put last.

\subsection{Weight}
\label{weight}

Each connected polygon of a mask may have a different weight.
In galaxy surveys,
the `weight' attached to a polygon is the completeness of the survey
in that polygon.
These weights must be supplied by the user.
If no weights are supplied, then the weight defaults to 1.

If the input polygons of a mask overlap,
then the policy adopted by the \mangle\ software
is to allow the weights of later polygons in polygon files
to override the weights of earlier polygons.
Thus for example,
to drill empty holes in a region,
one would put the polygons of the parent region first
(with weight 1, perhaps),
and follow them with polygons specifying the holes
(with weight 0).

There are three ways to apply weights to polygons.
The first way is simply to edit the polygon file or files
specifying the mask.
Attached to each polygon in a polygon file
is a line that includes a number for the weight;
one simply edits that number.

The second way is to specify weights in a file.
The \mangle\ software contains a facility
to read in these weights, and to apply them successively
to the polygons of a polygon file.
Suppose that you have your own software that returns a weight
given an angular position in a mask.
The \mangle\ software includes a utility (\S\ref{midpoint})
to create a file giving the angular position of a point
inside each polygon of a mask.
This file of angular positions becomes input to your own software,
which should create a file of weights,
which in turn can be fed back to \mangle.

The third way to apply weights to polygons
is to write a subroutine (in either fortran or c)
that returns a weight given an angular position,
and compile it into \mangle.
The \mangle\ software includes some template examples of how to do this.
%For example, the weights of the 2dF 100k survey
%from software written by Norberg \& Cole (?)
%are incorporated into \mangle.

In our experience, method two is the method of choice,
except in cases that are simple enough that method one suffices.

\subsection{Unify}
\label{unify}

The set of non-overlapping polygons that emerges from balkanizing and weighting
may be more complicated than necessary.
The \mangle\ software includes a facility for simplifying
the polygons of a mask, which we call unification.
Unification is not strictly necessary,
but it tidies things up,
and it can save subsequent operations,
such as harmonization (see \S\ref{harmonize}),
a lot of computer time. 

\unifyfig

Unification eliminates polygons with zero weight,
and does its best to merge polygons with the same weight.
The algorithm is to pass repeatedly through a set of polygons,
merging a pair of polygons wherever the pair can be merged
into a single polygon by eliminating a single abutting edge. 
Figure~\ref{unifyfig} illustrates an example of the unification procedure.

\nounifyfig

Unification does not necessarily accomplish the most efficient unification,
nor, as illustrated in Figure~\ref{nounifyfig},
is unification necessarily exhaustive.

\section{Harmonize}
\label{harmonize}

The \mangle\ software
contains several utilities that do various things with a mask.
One of the most important of these is a utility
to take the spherical harmonic transform of a mask,
a process we call harmonization.
In particular,
the area of a mask is proportional to the zeroth harmonic.
Computation of the area of a polygon is basic to several
of the \mangle\ algorithms.
For example, whether two polygons intersect
is determined by whether the area of their intersection is non-zero.

The method for computing the spherical harmonics of a mask
consisting of a union of polygons
is described in the Appendix of \citet{H93b}.
The algorithm is recursive and stable,
able to compute harmonics to machine precision to arbitrarily high order,
limited only by computer power and patience.
The recursion, as implemented in the \mangle\ software,
recovers correctly from underflow,
which can occur at large harmonic number $\el$.

While the recursive algorithm by itself is fast,
there is a numerical penalty to be
paid for allowing the polygons of a mask to have arbitrary shape:
the computation time for harmonics up to $\el$ increases as $\el^3$,
a pretty steep penalty when $\el$ is large.
The computation time is proportional to the number of edges of
the polygons of the mask,
and on a 750\,MHz Pentium III it takes 3 CPU minutes per edge
to compute harmonics up to $\el = 1000$. 

Thus the method is slow compared to fast algorithms
specially designed for regular pixelizations,
such as HEALPix
\citep{Gorski}.

The most time-consuming part of the computation is rotating the harmonics of
an edge from its natural frame of reference into the final frame of reference:
it is this rotation that takes $\el^3$ time.
The rotation is unnecessary if the edge is a
line of constant latitude in the final reference frame,
and the computation goes faster, as $\el^2$, in this case. 

Another acceleration is possible if two edges are related
by a rotation about the polar axis of the final frame.
Although computing the harmonics
$a^{(e)}_{\el m}$
of a single edge $e$ still takes $\el^3$ time,
the harmonics
$a^{(e^\prime)}_{\el m}$
of a second edge $e^\prime$ rotated right-handedly
by azimuthal angle $\phi$ from edge $e$ are
$a^{(e^\prime)}_{\el m} = \e^{- \im m \phi} a^{(e)}_{\el m}$,
which is fast to compute.

In practice,
the \mangle\ software currently implements the latter acceleration
only in the special case where (some of) the polygons of a mask are rectangles,
polygons bounded by lines of constant azimuth and elevation.
The acceleration applies only if at least two rectangles of the mask
have the same minimum and maximum elevation.
Two such rectangles need not be adjacent in the polygon file:
\mangle\ reorders the computation of polygons so as to take advantage
of acceleration where possible. 

\subsection{Harmonization algorithm}
\label{harmonizealgorithm}

For completeness, we give here an overview of the method
detailed by \citet{H93b}.

The spherical harmonic coefficients $\omega_{\el m}$ of a mask $\omega(\bn)$,
a function of angular direction $\bn$,
are defined by
\begin{equation}
\label{on}
  \omega ( {\bn} ) = \sum_{\el = 0}^{\infty}  \sum_{m = - \el}^\el
  \omega_{\el m} Y_{\el m} ( {\bn} )
\end{equation}
\begin{equation}
\label{olm}
  \omega_{\el m} =
  \int  \omega ( {\bn} ) Y_{\el m}^* ( {\bn} ) \, \dd o
\end{equation}
where $Y_{\el m}$ are the usual orthonormal spherical harmonics,
and $\dd o$ denotes an interval of solid angle about ${\bn}$.

The key mathematical trick is to convert the integral~(\ref{olm})
for $\omega_{\el m}$ from an integral
over the solid angle of the mask to an integral over its edges.
This is done by 
introducing the square $L^2$ of the angular momentum operator
${\bL} \equiv - \im \, {\bn} \times \upartial / \upartial {\bn}$
into the integrand
\begin{equation}
\label{olmL}
  \omega_{\el m} = 
  \int  \omega ( {\bn} )
   {L^2  \over \el ( \el + 1 )}
  Y_{\el m}^* ( {\bn} ) \, \dd o
  \quad
  ( \el \neq 0 )
\end{equation}
which is valid except for the monopole harmonic $\el = 0$,
dealt with below.
The Hermitian character of the angular momentum operator ${\bL}$
allows equation~(\ref{olmL}) to be rewritten
\begin{equation}
\label{olmp}
\omega_{\el m} =
  {1 \over \el ( \el + 1 )}
   \int  {\bL} \omega ( {\bn} )
   \cdot
  {\bL}^* Y_{\el m}^* ( {\bn} ) \, \dd o
\ .
\end{equation}
By assumption,
the mask is a sum over polygons $p$,
and $\omega(\bn)$ is a constant $\omega^{(p)}$ within each polygon.
It follows that ${\bL} \omega$ in equation~(\ref{olmp})
is a sum over polygons,
with the contribution from each polygon being a vector
whose magnitude is
$\omega^{(p)}$ times $\im$ times a Dirac delta-function,
and whose direction is along the boundary of the polygon,
winding right-handedly about the polygon.
Thus the integral~(\ref{olmp}) reduces to a sum of integrals
over the boundaries
$\upartial \omega^{(p)}$ of the polygons
\begin{equation}
\label{olmo}
  \omega_{\el m} =
  {\im \over \el ( \el + 1 )}
  \sum_{{\rm polygons} ~ p}
  \omega^{(p)}
  \oint_{\upartial \omega^{(p)}}
  \dd {\bn}
   \cdot
  {\bL}^* Y_{\el m}^* ( {\bn} )
\ .
\end{equation}
The boundary $\upartial \omega^{(p)}$ of a polygon is a set of edges,
so the integral in equation~(\ref{olmo})
becomes a sum of integrals over each edge of each polygon.
Thus a harmonic $\omega_{\el m}$
is a sum of contributions
$\omega^{(p)} a^{(e)}_{\el m}$
from each edge $e$ of each polygon $p$
\begin{equation}
  \omega_{\el m} =
  \sum_{{\rm polygons} ~p}
  \omega^{(p)}
  \sum_{{\rm edges} ~ e}
  a^{(e)}_{\el m}
\ .
\end{equation}
Hence the analytic problem reduces to that of determining the harmonics
$a^{(e)}_{\el m}$
of the edge $e$ of a polygon.
The problem is well suited to computation,
and could easily be parallelized if required
(currently, \mangle\ is not parallelized).

Stable recursive formulae for computing the harmonics
$a^{(e)}_{\el m}$
of a polygon edge are given by \citet{H93b}.
First,
the harmonics of an edge are computed
in a special frame of reference where the axis of the edge cap
is along the polar direction (the $\bz$ direction)
of the spherical harmonics.
The harmonics of the edge are then rotated into the actual frame of reference.
The most time consuming part of the computation is the second part,
the rotation, with the computational time going as $\el^3$.

The above derivation of the harmonics $\omega_{\el m}$ of the mask
fails for the monopole harmonic $\el = 0$,
for which equation~(\ref{olmL}) is invalid.
The monopole harmonic $\omega_{00}$ is
\begin{equation}
  \omega_{00} = {A \over ( 4\upi )^{1/2}}
\end{equation}
where $A$ is the weighted area of the mask,
a weighted sum of the areas $A^{(p)}$ of the polygons $p$
\begin{equation}
\label{A}
  A = \sum_{{\rm polygons} ~p} \omega^{(p)} A^{(p)}
  \ .
\end{equation}
The general formula for the area $A^{(p)}$ of a polygon $p$ is
\begin{equation}
  A^{(p)} =
    2\upi \chi^{(p)}
    -
    \sum_{{\rm edges} ~ e}
      \Delta \phi^{(e)} \cos \theta^{(e)}
    -
    \sum_{{\rm vertices} ~ v}
      \psi^{(v)}
\end{equation}
where $\chi^{(p)}$, an integer, is the Euler characteristic
(= faces minus edges plus vertices of any triangulation) of the polygon,
$\Delta \phi^{(e)}$ and $\theta^{(e)}$ are the lengths and polar angles of the
edges $e$ of the polygon,
and $\psi^{(v)}$ are the exterior angles ($\equiv$ $\upi$ minus interior angle)
at the vertices $v$ of the polygon.

The Euler characteristic $\chi^{(p)}$ of a polygon,
a topological quantity, is calculable topologically
--
it equals two plus the number of connected boundaries
minus twice the number of groups to which the boundaries belong
--
but in practice it is quicker to compute the Euler characteristic as follows.
First, in the trivial case that the polygon is the whole sky,
its area is $4\upi$.
Second, in the special case that the polygon is a single cap
with edge $e$, its area is $2\upi (1 - \cos \theta^{(e)})$.
Otherwise, the polygon is an intersection of two or more caps.
If the area of any one of the caps is less than or equal to $2\upi$,
then the area of the polygon must be less than $2\upi$,
so the Euler characteristic $\chi^{(p)}$ must take that integral value
which makes the area lie in the interval $[0, 2\upi)$.
This leaves the case where every one of the caps of the polygon has
area greater than $2\upi$.
The policy in this case is to introduce an extra cap which splits the polygon
into two parts, each of whose areas is less than $2\upi$;
the area of the polygon is then the sum of the areas of the two parts.
In practice,
polygons which are intersections of caps all of whose areas exceed $2\upi$
are fairly uncommon,
so the slow down involved in splitting these particular polygons is not great.
Moreover the splitting is necessary only for the monopole harmonic:
the higher order harmonics can be computed from the original polygon
without splitting it.

\subsection{Map}
\label{map}

The \mangle\ software contains a `map' utility to
reconstruct the mask at arbitrary points $\bn$ from spherical harmonics up to a
given maximum $\el_{\max}$
\begin{equation}
  \omega ( {\bn} ) =
  \sum_{\el = 0}^{\el_{\max}} \sum_{m = - \el}^\el
  \omega_{\el m} Y_{\el m} ( {\bn} )
\ .
\end{equation}
For example,
the bottom right panel of Figure~\ref{twoqzfig}
was generated using this utility.

\section{Cross and auto-correlations}
\label{correlation}

\subsection{Cross correlation $\langle DR \rangle$}

Another important feature provided in the suite of \mangle\ software
is a utility to compute precisely the angular cross-correlation
$\langle DR \rangle$
at given angular separation $\theta$
between given `Data' points
and `Random' points in the mask.
The angular integral is done analytically
and evaluated to machine precision,
rather than by Monte Carlo integration with Random points.

If the `Data' points are chosen randomly within the mask,
as in \S\ref{random},
then the cross-correlation becomes equivalent
to the angular autocorrelation
$\langle RR \rangle$,
for Random-Random,
at given angular separation between pairs of points in the mask.

The advantage of computing the
$\langle DR \rangle$
angular integral analytically
over the traditional Monte Carlo method
is that it eliminates unnecessary shot noise.
As discussed for example by
\citet{KSS},
this unnecessary shot noise can adversely affect the performance
of estimators of the correlation function at small scales.

In the case of the PSCz high latitude mask,
which balkanizes into 744 polygons,
it takes 5 CPU minutes per 1000 `Data' points to compute
$\langle DR \rangle$
at 1000 angular separations
with a 1.2\,GHz Pentium III.

\subsection{$\langle DR \rangle$ Algorithm}
\label{dr}

The contribution of a Data point at position $\bn$
to the angular correlation
$\langle DR \rangle$
at angular separation $\theta$
is a weighted sum over the polygons of the mask
of the azimuthal angle $\Delta \phi$ subtended within the polygon
by a circle of radius $\theta$ centred at $\bn$.
The correlation
$\langle DR \rangle$
at angular separation $\theta$
is an average over the contributions from each Data point.

The computation of
$\langle DR \rangle$
is done at a specified angular separation or set of separations $\theta$,
not, as is common with Monte Carlo integration,
over a bin of angular separations.
For a finite number of Data points,
and a mask consisting, as in the present paper, of a union of weighted polygons,
the correlation
$\langle DR \rangle$
is a continuous function of angular separation $\theta$,
except that,
if a data point $\bn$ happens to coincide with the axis of an edge of a polygon,
then the function will be discontinous at a separation $\theta$
equal to the polar angle of the said edge.
Furthermore,
the correlation $\langle DR \rangle$ is a differentiable function
of separation $\theta$ except at a finite, possibly large,
number of discontinuities where $\theta$ equals
the separation between a Data point $\bn$ and either a vertex
of a polygon or a point on the edge of a polygon where the separation
from $\bn$ is extremal.

Although
$\langle DR \rangle$,
as a function of separation $\theta$,
is thus typically discontinuous in the derivative,
and occasionally discontinuous in itself,
in practical galaxy surveys
$\langle DR \rangle$
tends to be relatively smooth,
especially when the number of Data points is large.
Thus in the practical case it is usually fine to sample
$\langle DR \rangle$
at a suitably large number of angular separations,
and to interpolate (linearly) on such a table.

The DR utility loops in turn through each Data point $\bn$,
so that two points take twice as much CPU time as one point.
For each Data point,
the DR utility attempts to accelerate the computation
with respect to the angular separations $\theta$ by first computing the
minimum and maximum angles between the point $\bn$ and each polygon in the mask.
The information about the minimum and maximum angles is used to decide whether
the circle about $\bn$ lies entirely outside or entirely inside a polygon,
in which case the (unweighted) angle $\Delta \phi$ subtended within the polygon
is zero or $2\upi$.
In practical cases the angle subtended is often zero for the great majority
of polygons of a mask, especially when the mask is composed of many polygons.
Since calculation of the subtended angle can be skipped if the angle is zero,
computation can be greatly speeded up.
Further acceleration comes from ordering the polygons in increasing order
of the minimum angle from the given point $\bn$ to each polygon.
This allows the computation to loop to the next value of $\theta$
as soon as it hits a polygon for which the subtended angle is zero,
rather than checking through large numbers of polygons that
all have zero subtended angle.

\subsection{Auto correlation $\langle RR \rangle$}

The angular auto-correlation
$\langle RR \rangle$
between pairs of points in a mask
can be computed in various ways.
The traditional method, as suggested by the Random-Random designation,
is to count pairs of random points.
However,
the traditional method is certainly not the most precise method,
and it may not be the most efficient, especially at small scales,
if large numbers of Random points are needed to reduce the shot noise
to a subdominant level.

An alternative method, mentioned above, is to compute
$\langle RR \rangle$
using the
$\langle DR \rangle$
algorithm with the `Data' points chosen randomly within the mask.
Although this method is subject to some shot noise,
the shot noise is liable to be substantially less
than that of the traditional method at small scales.

At the largest angular separations $\theta$,
a choice method is to compute
$\langle RR \rangle$
from its spherical harmonic expansion,
truncated at some suitably large harmonic $\el_{\max}$:
\begin{equation}
  \langle RR \rangle
  =
  2\upi
  \sum_{\el = 0}^{\el_{\max}} P_\el ( \cos \theta )
  \sum_{m = -\el}^{\el} \left| \omega_{\el m} \right|^2
\end{equation}
where $P_\el(x)$ is a Legendre polynomial,
which is accurate at angular scales $\gg \upi / \el_{\max}$.

The exact expression for
the autocorrelation
$\langle RR \rangle$
at angular separation $\theta$ is
\begin{equation}
  \langle RR \rangle
  =
  \int \!\!\! \int \omega(\bn_1) \omega(\bn_2) \delta_D(\bn_1 . \bn_2 - \cos \theta)
  \, \dd o_1 \dd o_2
  \ .
\end{equation}
For a mask of polygons as considered in this paper,
this double integral over solid angles
can be transformed into a double integral over the edges of the polygons
\citep{H93b}
\begin{eqnarray}
\label{RRexact}
\lefteqn{
  \langle RR \rangle
  =
  2\upi A
}
&&
\\
\nonumber
&&
  -
  \sum_{p_1, p_2}
  \omega^{(p_1)} \omega^{(p_2)}
  \oint_{\upartial \omega_1} \! \oint_{\upartial \omega_1} G(\cos\theta, \bn_1 . \bn_2)
  \, \dd \bn_1 . \dd \bn_2
\end{eqnarray}
where $A$ is the weighted area, equation~(\ref{A}), of the mask,
and $G$ is the Green's function of the scalar product $\bL_1 . \bL_2$
of angular momentum operators:
\begin{equation}
  G(x, x_{12}) =
  \left\{
  \begin{array}{ll}
    \displaystyle{
    \frac{1}{2}
    \ln
    \left[
       {(1 + x_{12}) (1 - x) \over (1 - x_{12}) (1 + x)}
    \right]
    }
    &
    (x_{12} \geq x)
    \\
    0
    &
    (x_{12} \leq x)
    \ .
  \end{array}
  \right.
\end{equation}
Unfortunately the integral~(\ref{RRexact}) cannot be solved analytically,
and we have not attempted to implement its numerical solution
in the \mangle\ software.
\citet{H93b}
gives a series expansion of the expression~(\ref{RRexact})
valid at small scales,
but the series expansion is liable to break down already at tiny
scales in the typically complex masks of modern surveys
(the coefficients of the series expansion change discontinuously
wherever the angular separation equals the distance between distinct vertices,
or more generally any extremal distance between pairs of edges),
making the series expansion of limited applicability.

\subsection{Generate random points inside a mask}
\label{random}

The \mangle\ software contains a utility for generating
random points inside a mask.
As indicated above,
this can be useful for example
in computing the angular auto-correlation
$\langle RR \rangle$
of pairs of points in the mask.

The algorithm, which is quite fast, is as follows:
\begin{enumerate}
\item
Select randomly a polygon in the mask,
with probability proportional to the product of the polygon's weight and area.
Lasso that polygon with a circle that is intended to be a tight fit,
but is not necessarily minimal.
Generate a point randomly within the circle,
test whether the point lies inside the polygon,
and keep the point if it does. 
\item
Iterate.
\end{enumerate}

A lasso is computed for a polygon as needed,
but is then recorded,
so that a lasso is computed only once for any polygon.
If the desired number of random points
exceeds the number of polygons in the mask,
then the computation starts by lassoing every
polygon in the mask. 

\section{Other Utilities}

The \mangle\ software
contains several other utilities,
described below.
%One utility not described below is a routine
%to rotate a set of azimuth-elevation points $\alpha$, $\beta$
%from one frame of reference into another;
%the routine is not described because it is essentially an
%interface to copyrighted {\sc starlink} routines,
%which cannot be included in the \mangle\ distribution.
%
%Future versions of the \mangle\ software may include other utilities
%not described here.

\subsection{Copying polygons into different formats}
\label{poly2poly}

One simple but oft-used utility is one that copies a polygon file or files
into another polygon file in a different format
-- see \S\ref{format} for an abbreviated description of the possible formats.
For example,
most of the figures in this paper were produced from points
generated by copying polygon files into graphics format.

The utility for copying polygon files has some switches
to copy polygons with weights only in some interval,
or with areas only in some interval.
This makes it easy for example
to discard polygons with small weights or small areas.

\subsection{The vertices and edges of a mask}
\label{vertices}

The \mangle\ software
contains routines that return the vertices of the polygons of a mask,
and to return the positions of points along the edges of a mask.
For example,
when a polygon file is copied into graphics format,
the copy utility invokes these routines.

A single polygon can have more than one connected boundary,
as illustrated in Figures~\ref{discpolfig}--\ref{fractalfig}
and \ref{chainfig}--\ref{wfig}.
Here the routines return the vertices and edge points
on distinct boundaries as distinct sets.

The routines to determine the distinct connected boundaries
of a polygon are used in stage 2 of the balkanization process,
\S\ref{balkanize}.

\subsection{Find points inside the polygons of a mask}
\label{midpoint}

The \mangle\ software contains a routine to find a point inside
each polygon of a mask.
The aim is to find a point that is squarely inside the polygon,
well away from its edges.
For example,
the weights attached to the balkanized set of polygons
shown in the lower left panel of Figure~\ref{twoqzfig}
were obtained by picking a point inside each polygon,
and evaluating the weight at that point
from the $1^\prime \times 1^\prime$ pixelized map provided by the 2QZ team.
The pixelization means that the map is reliable only away from the edges
of a polygon,
so it is important to pick the point squarely inside the polygon.

The algorithm finds one point for each distinct connected boundary
of a polygon.
If the polygon contains non-simply-connected parts,
as in Figures~\ref{eyefig}--\ref{fractalfig},
that means that the algorithm will return more points
than there are distinct connected parts of the polygon;
however, having more points than necessary is not a problem.

The algorithm to find a point inside a polygon is mildly paranoid.
%\footnote{
%We do not know a proof that the algorithm must always work,
%but we do not know a counterexample where it fails.
%}.
For each connected boundary of a polygon,
the algorithm first determines the barycentre of the
midpoints of the edges of the connected boundary.
This barycentre cannot be guaranteed to lie inside
the polygon, so it is not enough to stop here.
Instead, great circles are drawn from the barycentre
to each of the midpoints of the edges of the connected boundary.
On each of these great circles, the midpoint of the segment
of the great circle lying within the polygon,
with one end of the segment being the midpoint of an edge
of the connected boundary, is determined.
Each of these segment midpoints inside the polygon is tested to see how far
away it is from the nearest vertex or edge of the polygon,
including edges and vertices other than on the connected boundary.
The desired point inside the polygon is chosen to be
that segment midpoint which is furthest away from any vertex or edge.

\subsection{Find polygon(s) inside which a point lies}
\label{polyid}

The \mangle\ software contains a utility to find inside which polygon
or polygons of a polygon file a given point lies.
A point may lie inside zero polygons, or one polygon, or more than one polygon.
The utility takes no short cuts:
it tests all points against all polygons.
This can take time if there are large numbers of points
and large numbers of polygons. 

If the polygon file has been produced by balkanization,
then a point would normally lie inside at most one polygon.
However, a point at the edge of two abutting polygons
is considered to lie inside both polygons.

\section{Summary}
\label{summary}

The observing strategies of modern galaxy surveys
typically produce angular masks with complex boundaries
and variable completeness.
The purpose of this paper has been to set forward a scheme
that is able to deal accurately and efficiently with such angular masks,
and thereby to reduce both the labour and the chance for inadvertent error.
The fundamental idea is to resolve a mask into a union of
non-overlapping polygons each of whose edges is part of a circle
(not necessarily a great circle)
on the sphere.

The scheme has been implemented in a suite of software, \mangle,
downloadable from
\url{http://casa.colorado.edu/~ajsh/mangle/}.

The \mangle\ software includes several utilities
for accomplishing common tasks associated with angular masks of galaxy surveys.
% WHAT IS MANGLE GOOD FOR?
This includes generating random catalogues reflecting the angular selection function
(a tool employed in almost all galaxy survey analysis),
measuring the $\langle DR \rangle$ and $\langle RR \rangle$ angular integrals
(needed for estimating the correlation function),
and expanding the mask in spherical harmonics
(a key step in various techniques for measuring the power spectrum
and redshift space distortions).
The scheme was originally motivated by the nature of real angular masks
of real galaxy surveys,
and the underlying angular routines have been battle-tested over many years.
The full apparatus of the \mangle\ software has been used on
the 2dF survey by
\citet*{THX},
and on the SDSS survey by \citet{Tegmark}.

\section*{Acknowledgements}

AJSH was supported by
NASA ATP award NAG5-10763 and by NSF grant AST-0205981.
MT was supported by NSF grants AST-0071213 \& AST-0134999, NASA grants
NAG5-9194 \& NAG5-11099, and fellowships from the David and Lucile
Packard Foundation and the Cottrell Foundation.
We thank Yongzhong Xu for his contributions to an early
version of the balkanization software,
and Michael Blanton for suggestions.

\appendix

\section{Proof of simply-connectedness theorem}
\label{simplyconnected}

This Appendix proves the following theorem, invoked in \S\ref{balkanize2}
(see Table~\ref{deftable} for a definition of the term {\bf group}):

\noindent
{\bf
A connected part of a polygon is simply-connected
if and only if all the boundaries of the connected part belong to a single group.
}

First,
suppose that a polygon contains a region which is connected
but not simply-connected.
It is required to prove that the boundaries of this region
belong to at least two distinct groups.
By definition of simply-connectedness,
a continuous line can be drawn entirely inside the non-simply-connected region
such that one connected boundary of the region lies entirely
on one side of the line,
and another connected boundary of the region lies entirely
on the other side of the line.
In the polygon of Figure~\ref{eyefig}, for example,
such a line would circulate around the central boundary
while remaining inside the outer boundary.
The continuous line cannot intersect any of the circles forming the caps
of the polygon,
because if the line did intersect a circle,
then the line would be inside the polygon on one side of the intersection,
and outside the polygon on the other side of the intersection,
contradicting the assumption that the line lies entirely inside the polygon.
Hence the continuous line must partition the circles of the polygon
into two non-intersecting groups.
The two connected boundaries on either side of the continuous line
must therefore belong to two distinct groups,
as was to be demonstrated.

Conversely,
suppose that a polygon has boundaries
that belong to at least two distinct groups.
It is required to prove that the two groups delineate
a connected but non-simply-connected region of the polygon.
Consider the polygon, call it A,
formed by the intersection of all the caps of one group.
The polygon A so formed must consist of one or more simply-connected parts;
for if any connected part of A were not simply-connected,
then according the previous paragraph the boundaries of that part
would belong to different groups, contradicting the assumption
that the caps of A all belong to a single group.
Now consider similarly the polygon, call it B,
formed by the intersection of all the caps of a second group.
The boundaries of polygon B must lie entirely
inside one and only one of the simply-connected parts of A.
For certainly B must lie inside at least one part of polygon A,
since A must enclose the parent polygon,
and the boundary of B lies inside (at the border of) the parent polygon.
But the boundary of polygon B cannot lie in more than one part of A,
because if it did then,
because the circles of B are all in the same group and therefore
connected to each other,
there would be a path lying along the circles of B
traversing continuously from one part of A to another,
and therefore necessarily intersecting one of the boundaries of polygon A,
contradicting the assumption that the circles of A and B
belong to distinct groups that nowhere intersect.
Similarly,
the boundaries of polygon A must lie entirely
inside one and only one of the simply-connected parts of B.
This argument has identified two special boundaries:
a boundary of A that entirely encloses B;
and a boundary of B that entirely encloses A.
The region enclosed by these two special boundaries
delineates a region of the polygon that is connected but not simply-connected.
For consider a continuous line which is displaced slightly
off the special boundary of A, towards the special boundary of B.
By construction, the continuous line lies entirely inside both polygons A and B.
The continuous line cannot intersect any of the circles of A or B,
because if it did then the line would lie inside A (or B)
on one side of the intersection, and outside A (or B) on the other side,
contradicting the fact that the line lies entirely inside A and B.
The continuous line could possibly intersect circles belonging to
a third group of circles of the parent polygon.
However,
if the continuous line is displaced off the special boundary of A
by a sufficiently small amount that it does not encounter any third group,
then the continuous line will lie entirely inside the parent polygon.
This continous line forms a line inside the polygon that
cannot be shrunk continuously to a point,
so the polygon must contain a region that is connected but not simply-connected,
as was to be demonstrated.

This proves the theorem.

\end{document}